\documentclass[fleqn,usenatbib]{mnras}
\usepackage{newtxtext,newtxmath}
\usepackage{bm}
\usepackage[T1]{fontenc}
\usepackage{ae,aecompl}
\usepackage{graphicx}
\usepackage{amsmath}
\usepackage{amssymb}
\usepackage[makeroom]{cancel}
\usepackage{color}
\usepackage{ulem}

\def \cblu {\textcolor{blue}}

\newcommand{\tr}{$t_{\rm cool}/t_{\rm ff}$}

\definecolor{webgreen}{rgb}{0,.5,0}
\definecolor{webbrown}{rgb}{.6,0,0}

\title[Cool-Core Cycles and Phoenix]{Cool-Core Cycles and Phoenix}

\author[Prasad et al.]{
Deovrat Prasad$^{1}$\thanks{Email:deovratd@msu.edu},
Prateek Sharma,$^{2}$
Arif Babul,$^{3}$
G. Mark Voit,$^{1}$
and Brian W. O'Shea$^{1,4}$\\
$^{1}$ Department of Physics and Astronomy, Michigan State University, MI, USA \\
$^{2}$ Department of Physics, Indian Institute of Science, Bangalore, India \\
$^{3}$ Physics and Astronomy, University of Victoria, Victoria, Canada \\
$^{4}$ Department of Computational Mathematics, Science, and Engineering, National Superconducting Cyclotron Laboratory,\\
and Institute for Cyber-Enabled Research, Michigan State University, MI, USA
}
\date{}

\begin{document}
\label{firstpage}
\pagerange{\pageref{firstpage}--\pageref{lastpage}}
\maketitle
\begin{abstract}
Recent observations show that the star formation rate (SFR) in the {\it Phoenix} cluster's central galaxy is $\sim 500$ M$_\odot$ yr$^{-1}$. Even though {\it Phoenix} is a massive cluster ($M_{200} \approx 2.0\times 10^{15}$ M$_\odot$; $z\approx 0.6$) such a high central SFR is not expected in a scenario in which feedback from an active galactic nucleus (AGN) maintains the intracluster medium (ICM) in a state of rough thermal balance. It has been argued that either AGN feedback saturates in very massive clusters or the central supermassive black hole (SMBH) is too small to produce enough kinetic feedback and hence is unable to quench the catastrophic cooling. In this work, we present an alternate scenario wherein intense short-lived cooling and star formation phases followed by strong AGN outbursts are part of the AGN feedback loop. Using results from a 3D hydrodynamic simulation of a standard cool-core cluster ($M_{200}\sim 7\times10^{14}$ M$_\odot$; $z=0$), scaled to account for differences in mass and redshift,
we argue that {\it Phoenix} is at the end of a cooling phase in which an AGN outburst has begun but has not yet arrested core cooling. This state of high cooling rate and star formation is expected to last for $\lesssim$ 100 Myr in {\it Phoenix}.     
\end{abstract}
\begin{keywords}
Galaxy Clusters, Cooling flow, Black Hole, AGN feedback, Intra-cluster medium 
\end{keywords}
\section{Introduction}\label{sec:introduction}
Recent deep multi-wavelength observations of the {\it Phoenix} cluster by \citet{McDonald2014,McDonald2019,McDonald2019a} have detected a star formation rate of $\sim 500$ M$_\odot$ yr$^{-1}$ ($z\approx0.6$) in the central galaxy ($r<50$ kpc). Such a high star formation rate is not expected in the standard AGN feedback scenario, even in a massive galaxy cluster like {\it Phoenix} ($M_{200}\approx 1.8\times10^{15}$ M$_\odot$).  However, it is consistent with the condensation rate of a pure cooling flow in the intracluster medium (ICM). {\it Phoenix} is therefore unlike CL09104, another massive galaxy cluster in which the high star formation rate ($70-200$ M$_\odot$ yr$^{-1}$) is due to an ongoing merger (\citealt{ewan2012}). Given that the cooling time, $t_{\rm cool} \equiv 3nk_BT/[2 n_e n_i\Lambda]$, in the central $r<20$ kpc of {\it Phoenix} is an order of magnitude shorter than in any other observed galaxy cluster, \citet{McDonald2019} have speculated that {\it Phoenix} has experienced an evolutionary pathway different from other cool core clusters. 

The intracluster medium (ICM) is expected to become multiphase whenever min($t_{\rm cool}/t_{\rm ff}$) $\lesssim 10$ (with $t_{\rm ff} \equiv \sqrt{2r/g}$), leading to star formation and AGN outbursts (\citealt{mccourt12,sharma12,Prasad15}). Most cool core clusters are observed to be in a state of rough thermal balance, with min($t_{\rm cool}/t_{\rm ff}$) between $10-20$ (\citealt{voit15N, voit15E, Hogan2017, Pulido2018}).  Few have \tr~$< 10$, but {\it Phoenix} has \tr~$\sim 1$ at $r\approx3$ kpc (\citealt{McDonald2019}).  None with \tr~$< 5$ were previously known, making {\it Phoenix} a significant outlier.

{\it Phoenix} has a large X-ray luminosity ($\approx 4.7\times10^{45}$ erg s$^{-1}$; \citealt{ueda2013}) and a kinetic feedback power $\sim 10^{46}$ erg s$^{-1}$ inferred from the cavities the AGN has produced, assuming the AGN has shut off and the cavities are rising buoyantly (\citealt{Russell2017,McDonald2019}).  The cavity power is therefore comparable to the X-ray luminosity, but it has been argued that the mechanical jet power may be saturated and unable to quench the cooling flow, perhaps because of a central black hole mass that is small compared to the halo mass (\citealt{McDonald2019}).  

In this work, we argue that {\it Phoenix} is not fundamentally different from other cool-core clusters and does not necessarily represent a failure of kinetic SMBH feedback. We suspect instead that {\it Phoenix} is near the end of a short-lived cooling state, with AGN feedback just turning back on after a $\ll 1$ Gyr starburst\footnote{Note added during pre-publication: The age  of the cavities in {\it Phoenix} from  the latest observations is estimated to be < 10 Myr (\citealt{akahouri2020})}. Given that the fraction of young stellar population is small, the current starburst cannot be long lived.  Similar events occur in the fiducial simulation of \citet{Prasad2018},\footnote{The 3D hydrodynamic NFW+BCG run without stellar depletion.} in which a strong AGN outburst follows each phase of core cooling. This transitional state lasts for only $\sim 50-100$ Myr, after which the AGN heats and disperses the core to a state more typical of cool core clusters.  In simulations, such cooling phases are seen across a range of halo masses but are most frequent in the most massive halos, given a fixed efficiency for converting mass accretion into kinetic feedback power (\citealt{Prasad15,li15,Prasad2018}).  Cooling phases are more frequent in very massive halos because even strong feedback is unable to push low-entropy gas far from the center.  

\section{Cooling and Initial AGN Heating Phase}
\label{sec:CHP}
In the simulation discussed here, the ICM was initialised in hydrostatic equilibrium in a gravitational potential given by the sum of a Navarro-Frenk-White (NFW) potential for the dark matter halo and a singular isothermal sphere (SIS) potential for the central brightest cluster galaxy (BCG).  The fiducial run initially has $\min(t_{\rm cool}/t_{\rm ff}) \approx 7$ at $r\approx$10 kpc and a minimum cooling time $\approx 200$ Myr. An accretion rate ($\dot{M}_{\rm acc}$) calculated at 0.5 kpc determines the power $\epsilon \dot{M}_{\rm acc} c^2$ of bipolar jets of kinetic feedback injected at small radii (for details see \citealt{Prasad15,Prasad2018}).
Accretion and jet injection are minimal at first because the galaxy cluster is initialized in hydrostatic equilibrium. After a core cooling time ($\approx 200$ Myr) the cooling-flow accretion rate rises, resulting in enhanced AGN jet power.  AGN outbursts then heat the core, push gas around, and raise $t_{\rm cool}/t_{\rm ff}$, keeping the average $\dot{M}_{\rm acc}$ well below the cooling-flow value. Throughout its lifetime, the cool core undergoes multiple radiative cooling and AGN heating cycles. 
\begin{figure}
    \includegraphics[width=0.47\textwidth]{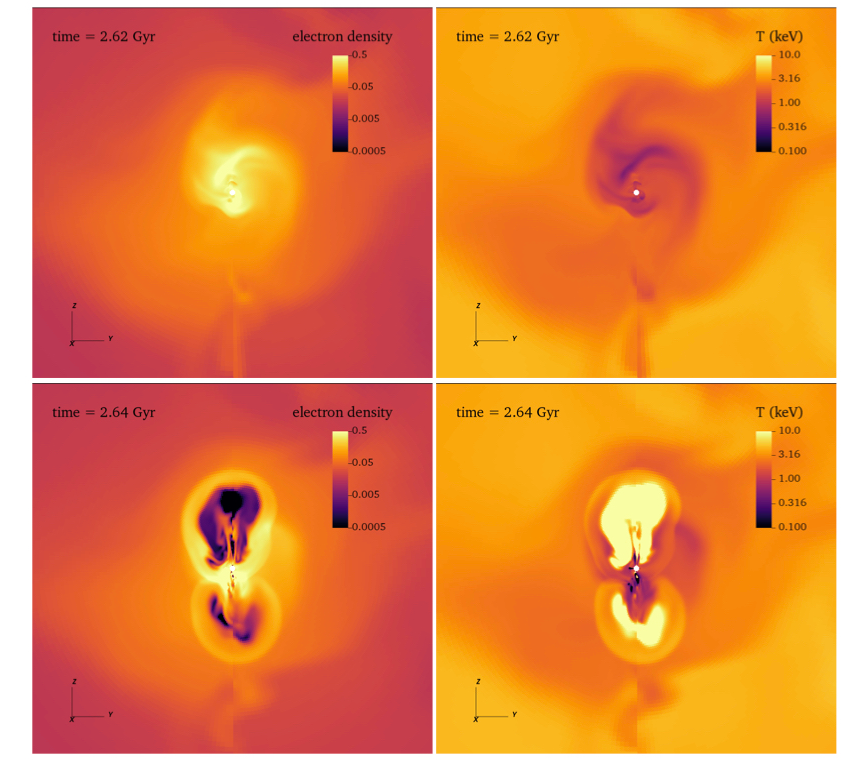}
    \caption{Electron number density ($n_e$) in cm$^{-3}$ (upper panels) and temperature ($T$) in keV (lower panels) in 60 kpc $\times$ 60 kpc slices within the plane of jet injection.  Two epochs are shown, one near the end of the cooling phase \cblu{(at 2.62 Gyr, top panels)} and the other at the beginning of the AGN outburst phase (at 2.64 Gyr, bottom panels). The cooling phase exhibits filamentary structures condensing out in the core ($r \lesssim 30$ kpc). During the initial AGN outburst phase, high density shocked regions can be seen around the cavities inflated by AGN jets.}
    \label{fig:den}
\end{figure}

Figure \ref{fig:den} shows density (left panels) and temperature (right panels) slices along the jet axis during a cooling phase at $t=2.62$ Gyr (top panels) and the subsequent AGN outburst at $t =2.64$ Gyr (bottom panels). Those epochs correspond to a strong starburst before AGN feedback couples to the ICM and quenches star formation (see the right panel in Fig. 4 of \citealt{Prasad2018}). 

\begin{figure}
    \includegraphics[width=0.5\textwidth]{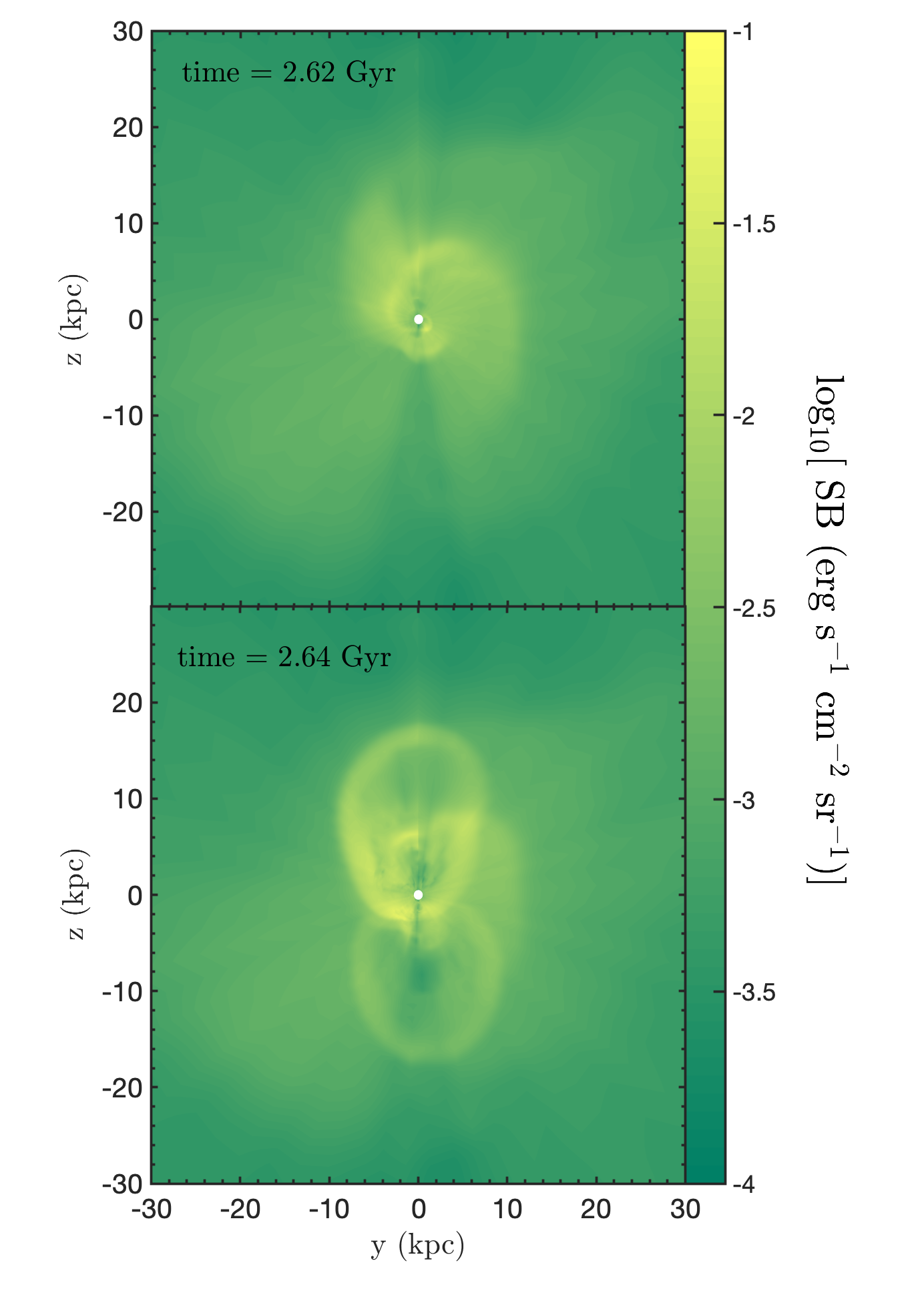}
    \caption{X-ray ($0.5-10$ keV) surface brightness maps (projected perpendicular to jet injection) corresponding to the cooling phase at $t=2.62$ Gyr (upper panel) and the beginning of AGN outburst phase at $t=2.64$ Gyr (lower panel). 
    Cooling phase shows a high X-ray surface brightness in the central ($r\sim 10$ kpc) regions with an elevated surface brightness extending well beyond 30 kpc.
    Notice the sharp jump in surface brightness due to the shocked gas around the cavity inflated by jets. }
    \label{fig:SB}
\end{figure}

Figure \ref{fig:SB} shows projected surface brightness maps of the X-ray gas (0.5-10 keV) at the same two epochs as Figure \ref{fig:den}. During the cooling phase the X-ray surface brightness is relatively smooth within $r \sim 10$ kpc (upper panel). But as AGN jets begin to make their way out of the core, pronounced surface brightness features appear in the form of bright rims around the jet-driven cavities (lower panel).

Morphologically, {\it Phoenix} appears to be in a state similar to the \cblu{$t=2.64$} Gyr snapshots (compare our Figs. \ref{fig:den} \& \ref{fig:SB} with Figs. 1, 2, and 5 in \citealt{McDonald2019}). Current star formation, as traced by UV and [OII], appears enhanced at the periphery of the X-ray cavity created by the AGN outburst. The bipolar X-ray cavities observed in {\it Phoenix} are attached to the center and have a major axis of $\sim 25~{\rm kpc}$, rather similar to (but somewhat bigger than) those in our simulations at \cblu{2.64} Gyr.

Cooling phases followed by AGN outbursts are expected to be more frequent in more massive halos for a fixed accretion efficiency because of the deeper potential well (for details, see Section 3.2.2 in \citealt{Prasad15}). This relationship suggests that the likelihood of observing a massive halo in a cooling state may be much greater than for a lower mass halo, possibly explaining why the first detection of a galaxy cluster at the end of a cooling phase happens to be a very massive cluster. 

\section{Thermodynamic Profiles}
\label{sec:one-d}
{\it Phoenix} exhibits profiles of entropy, cooling time, and $t_{\rm cool}/t_{\rm ff}$ in its core, $r<0.02\times r_{200}$ ($\sim 50$ kpc), that are unique compared to any other observed cluster. However,a comparison of the {\it Phoenix} cluster's scaled\footnote{We assume WMAP9 (\citealt{Bennett13}) cosmology to compute the characteristic parameters used to rescale the profiles.} thermodynamic profiles with those from our simulation at epochs corresponding to
the transitions between cooling and AGN outburst phases in our fiducial run reveals close similarities.
\begin{figure*}
 \includegraphics[width=1.0\textwidth]{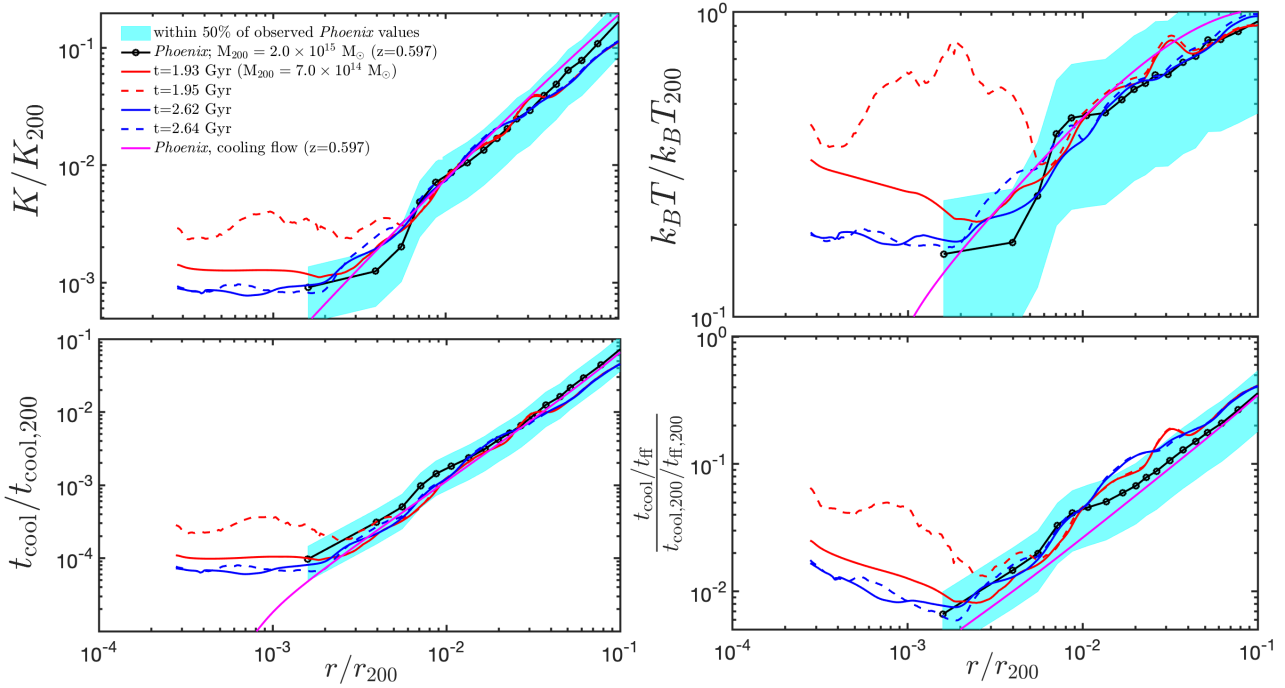}
 \caption{Angle-averaged, emissivity-weighted (for $0.5-10$ keV) and scaled entropy (top left panel), $T_{\rm keV}$ (top right panel), cooling time ($t_{\rm cool}$, lower left panel) and $t_{\rm cool}/t_{\rm ff}$ (lower right panel) radial profiles of four epochs from our simulation (red and blue curves).  At $t=1.93$ and $2.62$ Gyr (solid curves), the simulated cluster is in the intense cooling phase of the cooling cycle while $t=1.95$ and $2.64$ Gyr (dashed curves) marks the end of the cooling cycle and the beginning of the jet outburst phase. The observed profiles for {\it Phoenix} (\citealt{McDonald2019}) and a pure cooling flow for a Phoenix mass model based on read-off from Figure 10 \& 11 of \citet{McDonald2019} are the solid black line with circle markers and solid magenta line respectively . The cyan shaded region represent the $50\%$ scatter around the observed {\it Phoenix} values. The profiles are scaled to their characteristic values indexed to $r_{200}(M_{200},z)$ (assuming WMAP cosmology): $K_{200}(z) \equiv T_{\rm keV, 200}(z) \,  n_{e,200}(z)^{-2/3}$; $T_{200}(z) \equiv  G M_{200}\mu m_p  / [2 r_{200}(z) k_B]$; $t_{\rm cool, 200}(z) \equiv 1.5 \mu_e \mu_i m_p k_B T_{200}(z) / [ 200 \mu f_b \rho_{\rm cr}(z) \Lambda(T_{200}) ]$; and $t_{\rm ff,200} \equiv \sqrt{2r_{200}(z)/g_{200}(z)}$, where $g_{200}(z)= GM_{200}/r_{200}^2$.  Here, the symbols have their usual meaning. Once the differences in mass and redshift between the {\it Phoenix} cluster and our simulated cluster are accounted for through rescaling, simulation profiles at $t=2.62$ Gyr are in excellent agreement with the {\it Phoenix} result.}
 \label{fig:cp_1d}
\end{figure*}

\begin{figure}
 \includegraphics[width=0.5\textwidth]{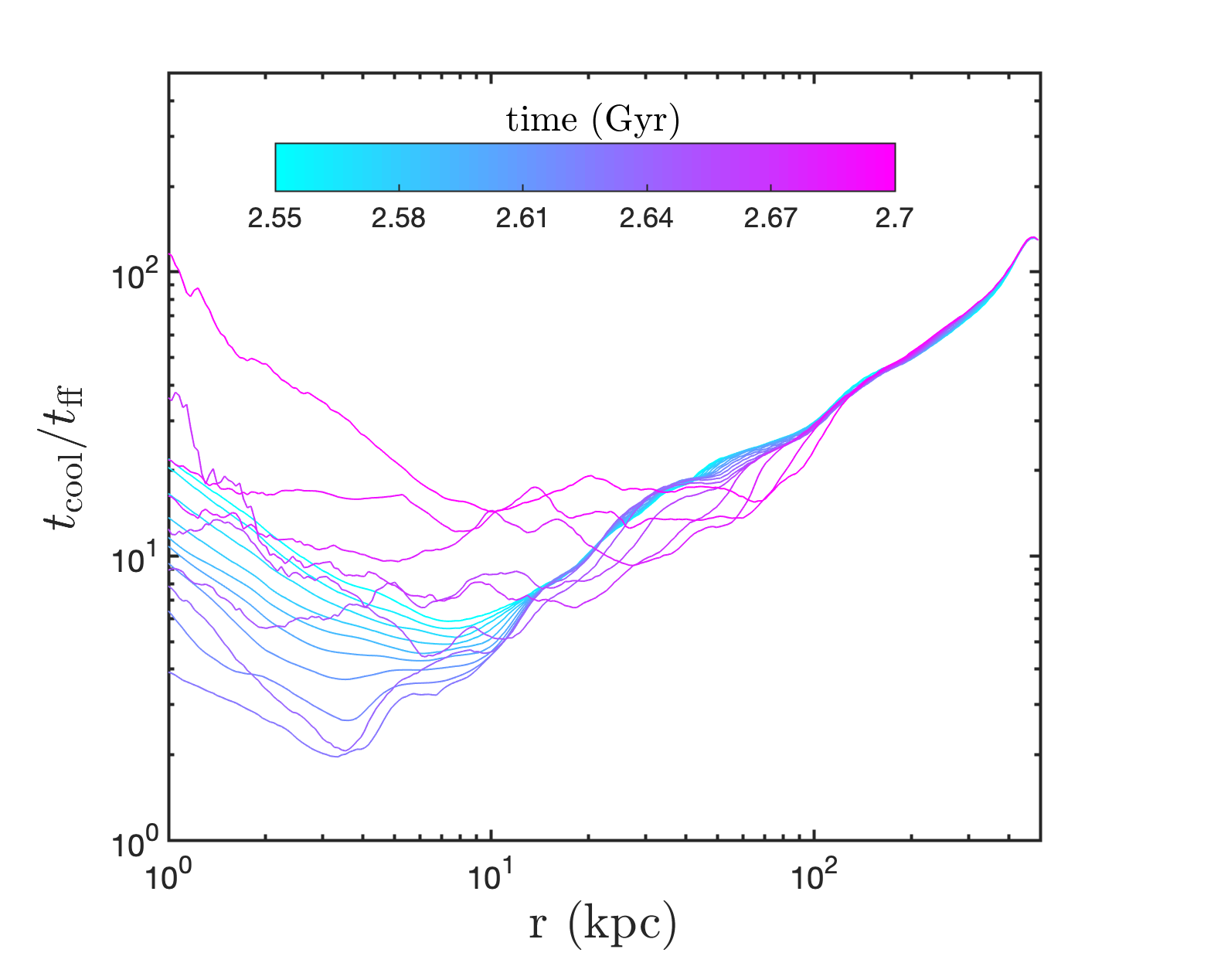}
 \caption{Angled-averaged emissivity-weighted (for $0.5-10$ keV) $t_{\rm cool}/t_{\rm ff}$ radial profiles of the X-ray emitting gas ($0.5-10$ keV) from $t=2.55-2.70$ Gyr for our fiducial run. The colours of the lines represent time as shown in the colour bar. The plot shows the variation in $t_{\rm cool}/t_{\rm ff}$ over 150 Myr as the cluster goes through intense cooling and rapid AGN heating phases.}
  \label{fig:tctf_color}
\end{figure}

The top left panel in Figure \ref{fig:cp_1d} shows the azimuthally-averaged, emissivity-weighted scaled entropy profile
(\citealt{voit05,mcc08,Kaiser86}) of X-ray gas (0.5-10 keV) for two different {\it Phoenix}-like cooling cycles, the $t=1.85-2.0$ Gyr (red lines) and the $t=2.55-2.7$ Gyr cycles (blue lines). The solid lines show the entropy profiles during the cooling phase and the dot-dashed lines show the profiles at the start of the corresponding AGN outburst phases. Entropy in the central $r\lesssim 0.005r_{200}$ ($\sim 10$ kpc) falls below 10 keV cm$^2$ during the cooling phase and in the case of the second cycle, remains below this value even 20 Myr after the onset of AGN feedback, as the jets drill their way out of the core. The scaled entropy profile of {\it Phoenix} (solid black line with circles) shows strikingly similar behaviour.  In the discussion below, we focus on the second cycle, which as discussed in \S2, more closely resembles {\it Phoenix } from a morphological perspective.

During the intense cooling phase, the cooling time (bottom left panel) in the central $r\lesssim 0.005 r_{200}$ ($\lesssim 10$ kpc) region of our simulated cluster falls to $\sim 10^{-4}t_{\rm cool} (r_{200})$ ($\sim 10s$ of Myr), and remains below 100 Myr in the core \cblu{($r<0.005 r_{200}$ kpc)} during the early stages of the AGN outburst phase. While the magnitude of {\it Phoenix} cluster's
cooling time profile, $t_{\rm cool}(r)$, is a factor of 2 lower at all radii, the scaled $t_{\rm cool}$ curve is remarkably similar to our simulation results from $r\sim 0.1 r_{200}$ ($\sim 200$ kpc)  down to $0.001 r_{200}$ ($\sim 2-3$ kpc). 

The scaled emissivity-weighted temperature plot (top right panel, Figure \ref{fig:cp_1d}) shows that the temperature profile of {\it Phoenix} over the radial range $r < 0.1r_{200}$ is likewise consistent with the cooling phase of our simulated cool core cluster during the second cycle.
The individual wiggles in the temperature profile are due the presence of AGN cavities. 

The bottom right panel in Figure \ref{fig:cp_1d} shows the scaled $t_{\rm cool}/t_{\rm ff}$ profiles from our simulation.
Before the AGN turns on, $t_{\rm cool}/t_{\rm ff}$ reaches its minimum value of $0.01\times t_{\rm cool,200}/t_{\rm ff,200}$ ($\sim 2-3$) at $r\sim 0.002 r_{200}$ ($\sim 5$) kpc and rises approximately linearly toward larger radii.  {\it Phoenix} similarly exhibits an approximately linear rise in \tr from $0.001~r_{200}$ ($\sim 2-3$ kpc) out to large radii. The simulation and {\it Phoenix} scaled profiles are in excellent agreement.

Figure \ref{fig:tctf_color} shows how the (unscaled) $t_{\rm cool}/t_{\rm ff}$ profile from our simulation
varies during a 150 Myr period that includes the intense cooling and the subsequent AGN outburst phases at $t=2.55-2.7$ Gyr. Line colours represent time as shown in the colour bar. During the initial cooling phase, $t_{\rm cool}/t_{\rm ff}$ declines in the central $r<30$ kpc, and the plasma becomes unstable to multiphase condensation, producing cold gas clumps that raise the core density and fuel a strong AGN outburst. These profiles show the first signs of AGN heating at 2.64 Gyr within $r<3$ kpc. At 2.65 Gyr the effect of AGN heating can be seen extending to $r<10$ kpc, although min$(t_{\rm cool}/t_{\rm ff})$ is still $\approx 5$. By 2.68 Gyr, the AGN outburst has driven the core to min($t_{\rm cool}/t_{\rm ff}$)$\approx 10$, and has driven it to $\approx 20$ by 2.7 Gyr.       
The figure shows that $\min(t_{\rm cool}/t_{\rm ff}) \lesssim 5$ is maintained for the first $\lesssim20$ Myr of the active AGN phase. Similar behaviour is seen towards the end of an earlier cooling cycle at
$t=1.85-2.0$ Gyr (see Fig. \ref{fig:cp_1d}, as well as the right panel of Fig. 4 in \citealt{Prasad2018}). 

During the $t=1.85-2.0$ Gyr cycle, the cooling rate in the central 10 kpc is $\sim 200$ M$_\odot$ yr$^{-1}$ (about a third of the steady cooling flow rate), leading to accumulation of $\gtrsim 5\times10^9$ M$_\odot$ of cold gas ($T<10^4$ K) by 1.95 Gyr.
The cooling rate dies down after this as the jets heat up the core, partly evaporating the cold gas leading to a slight decrease in the total cold gas.  During the second cooling cycle at $t=2.55-2.7$ Gyr, we observe a similar behaviour as well as enhanced cooling mostly along the periphery of the cavities at 2.62-2.66 Gyr as the jet cavities inflated.  The latter leads to an 
accumulation of about $2\times10^9$ M$_\odot$ of cold gas ($T<10^4$ K) and a dropout of $> 10^9$ M$_\odot$ of intermediate temperature ($10^4 K < T < 5\times10^6$) gas.  
This is an example of positive feedback \citealt{combes17}).
Cold gas mass starts to decline after 2.66 Gyr as the shocks and mixing gradually heat up the core.

For comparison, Figure \ref{fig:cp_1d} also shows the scaled profiles (magenta solid lines)
for a steady cooling flow applied to the mass model of {\it Phoenix} ($M_{200}=2.5 \times 10^{15} M_\odot$, $c_{200}=10$, $V_c=250$ km s$^{-1}$) and appropriate boundary conditions at 500 kpc ($n_e=10^{-3}$ cm$^{-3}$, $K=1057$ keV cm$^{2}$). 
The steady cooling flow model is similar to the observed profiles (\citealt{McDonald2019}; see also \citealt{Stern2019}) but so are the profiles from our simulations (once the mass and redshift differences are taken into account) with feedback in the cooling phase. Thus, similarity to a cooling flow solution does not necessarily imply an absence of feedback heating. 

\section{Observational and Theoretical Limitations}
\label{limitations}
While the similarities between {\it Phoenix} and the cooling phases of our simulated cluster are encouraging, observations of other cool core clusters show that few have $1 \lesssim \min(t_{\rm cool}/t_{\rm ff}) \lesssim 10$.  However, the temporal distribution of min($t_{\rm cool}/t_{\rm ff}$)
in our simulation (see Figure 11 of \citealt{Prasad2018}) suggests that systems with min($t_{\rm cool}/t_{\rm ff}) \lesssim 5$ should be present. 
 
One possibility is that {\it Phoenix}-like cooling episodes are more common in our idealized non-cosmological simulations than in real clusters.  For example, our idealized simulations do not include cosmological structure formation, which can produce dynamical perturbations capable of promoting multiphase condensation at higher levels of \tr ~(\citealt{Choudhury2019}), preventing lower levels from being reached.  Consequently, the temporal distribution of $\min (t_{\rm cool}/t_{\rm ff})$ for a single, idealized, non-cosmological simulated cluster might not be representative of a heterogenous cluster population with very different
accretion histories.  Additionally, our
simulations (\citealt{Prasad2018}) does not include magnetic fields or anisotropic thermal conduction, which may also lower the incidence of very small \tr.  All of the above effects will result in less {\it Phoenix}-like states than in our simulations.

Another possibility is that low values of $\min(t_{\rm cool}/t_{\rm ff})$ like the one observed in {\it Phoenix} can be detected only in deep observations that provide a large enough photon count at $r < 10$~kpc. In \cite{McDonald2019}, the large photon count  allowed accurate removal of the AGN X-ray point source, revealing sharply peaked diffuse emission indicative of unusually low entropy gas. Earlier observations without accurate point-source removal estimated the central temperature of {\it Phoenix} to be $\approx 7$ keV and the central entropy to be $>10$ keV cm$^2$ (see Phoenix [SPT-CLJ2344-4243] in Figs. 3, 5 of \citealt{McDonald2019a}), significantly greater than the values obtained by \citealt{McDonald2019}. 
\begin{figure}
 \includegraphics[width=0.45\textwidth]{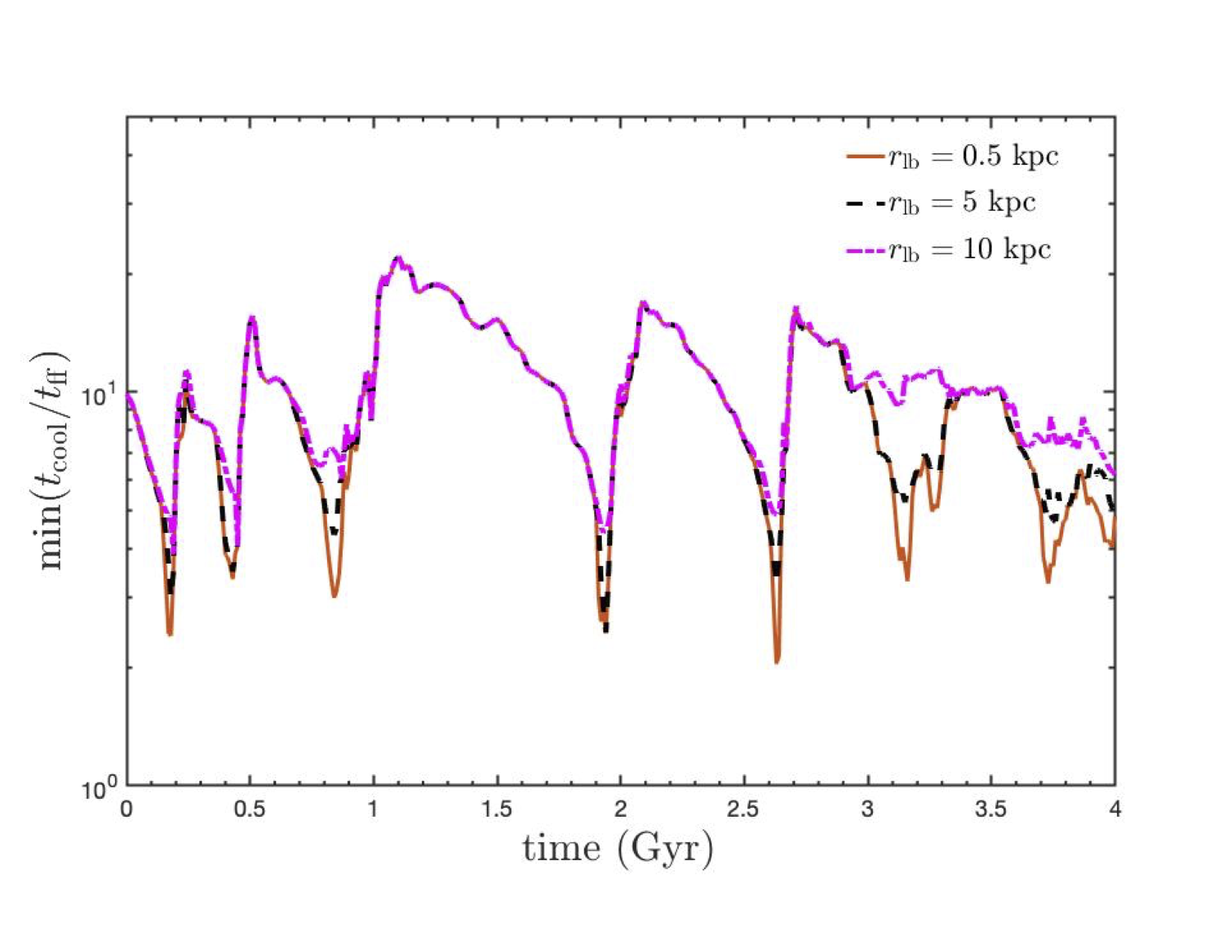}
 \caption{The variation of min($t_{\rm cool}/t_{\rm ff}$) with time for different inner cut-off radius for our fiducial run.
 This plot shows the importance of high spatial resolution to obtain the correct value of min($t_{\rm cool}/t_{\rm ff}$) in cool core clusters. This important parameter can be grossly overestimated if the spatial resolution is $\gtrsim 10$ kpc. }
  \label{fig:mintctf}
\end{figure}

A greater photon count also enables more finely grained radial binning, which improves the spatial resolution achievable with deprojection. Figure \ref{fig:mintctf} tries to quantify the effects of spatial resolution on observational estimates of min($t_{\rm cool}/t_{\rm ff}$). It shows how min($t_{\rm cool}/t_{\rm ff}$) varies with time in our fiducial simulation for different inner cut-off radii. The dot-dashed magenta line shows min($t_{\rm cool}/t_{\rm ff}$) for $r>10$ kpc, the dashed black line for $r>5$ kpc and the solid orange line for $r>0.5$ kpc. At high resolution, the dips in min($t_{\rm cool}/t_{\rm ff}$) are more pronounced while the larger values remain unchanged because the region with $t_{\rm cool}/t_{\rm ff} \lesssim 5$ is relatively small. In contrast, the minimum $t_{\rm cool}/t_{\rm ff}$ achieved in a heating phase is at $r\gtrsim10$ kpc (see Fig. \ref{fig:tctf_color}). Observations with high angular resolution (enabled by a high photon count) are thus needed to accurately extract min($t_{\rm cool}/t_{\rm ff}$) for clusters in a cooling phase.  Some of the deep observations from \citealt{babyk2018} do find min(\tr) $\approx 5$ in elliptical galaxies. However, it remains possible that the lowest observed values of min($t_{\rm cool}/t_{\rm ff}$) are biased low because of noise. Similar interpretations have been made for the low min($t_{\rm cool}/t_{\rm ff}$) clusters in \citealt{Pulido2018}, in which nearly half of the clusters with confirmed CO detection lie between 7-12 with a distribution peaking at 10-12 (see Figure 11, \citealt{Prasad2018}). Making more definitive comparisons with observations will require deprojection of mock observations of simulated clusters, so that the values of min($t_{\rm cool}/t_{\rm ff}$) derived from the simulations can more accurately reflect what observers actually measure.  However, such an analysis is beyond the scope of this short paper.

\section{Discussion}
\label{disc}
We find that many of the unique features of the {\it Phoenix} cluster core are consistent with a numerical simulation of kinetic AGN feedback fueled by cold gas accretion. They do not necessarily signal a different evolutionary pattern for {\it Phoenix}, but rather motivate higher sensitivity observations of the most centrally peaked systems to assess how compatible the observed distribution of min($t_{\rm cool}/t_{\rm ff}$) is with the simulations. 

{\it Phoenix} has a high current SFR but its fraction of young stars indicates that it has not been forming stars at that rate for very long. The observed amount of molecular gas ($M_{\rm cold}>10^{10}$ M$_\odot$) most likely has originated in gas cooling from the surrounding hot atmosphere over $50-120$ Myr (\citealt{Russell2017}). Pure cooling for a few hundred Myr (Figure \ref{fig:tctf_color}) is enough to make the cluster look like a pure cooling flow, consistent with our interpretation of a short cooling cycle. Moreover, pure cooling in {\it Phoenix} for $\sim 1$ Gyr would create a cold gas reservoir of $10^{12}~M_\odot$, more than an order of magnitude larger than what is observed (\citealt{Russell2017}).

We cannot rule out that {\it Phoenix} has a black hole with an unusually low mass for its halo mass and therefore cannot produce enough kinetic feedback, as suggested by \citet{McDonald2019}.  However, if the black hole is indeed ineffectual, the feedback failure must be a very recent phenomenon otherwise, for the reasons noted above, 
we would expect a higher fraction of young stars and a much higher reservoir of cold gas than observed. 
\citealt{McDonald2012} used the scaling relations between spheroid stellar mass and central black hole mass (\citealt{bennert11}) to estimate a mass $\sim 1.8\times10^{10}$ M$_\odot$ for the SMBH, which would make it one of the most massive in the universe. Owing to its large Eddington accretion rate, such a massive SMBH can provide sufficient kinetic feedback through accretion at $10^{-2}$ times the Eddington rate.  The cavity power inferred in {\it Phoenix} may therefore be an underestimate, given our hypothesis that the AGN is in the initial stages of an outburst and may still be rapidly inflating its cavities.

\section{Conclusions}
\label{sec:conc}
The principal conclusions of this work are as follows:
\begin{itemize}
    \item Surface brightness snapshots of our cluster simulations, along with the corresponding radial profiles of density, entropy, temperature, and \tr\ during transitions from a cooling phase to an AGN outburst, are similar to observations of {\it Phoenix} (although the halo mass of the simulated cluster is a factor of few smaller).
    \item {\it Phoenix} is not necessarily fundamentally different from other cool-core clusters. We may have  caught it at the end of a cooling state, just as AGN feedback is turning back on. The thermodynamic and various timescale profiles from our simulation is in very good agreement with {\it Phoenix} profiles once the profiles are scaled to account for mass and redshift differences.
    \item {\it Phoenix} does not necessarily signal the failure of kinetic SMBH feedback. Similar phases may also be seen in lower mass halos.  However, the cooling phases in simulated systems are more frequent in more massive halos, given a fixed efficiency parameter. 
    \item Deeper observations with a higher photon count within $r<10$ kpc of centrally peaked galaxy clusters may be required to measure min($t_{\rm cool}/t_{\rm ff}$) accurately, since $t_{\rm cool}/t_{\rm ff}$ during a cooling phase is close to the minimum value only over a small range in radius ($r \ll 10$ kpc).
\end{itemize}

\section*{acknowledgments}
DP is supported by {\it Chandra} theory grant no. TM8-19006X (G. M. Voit as PI) and NSF grant no. AST-1517908 (B.W.O'Shea as PI). BWO acknowledges further support from NASA ATP grant NNX15AP39G.
A.B. acknowledges support from NSERC through the Discovery Grant program.
P.S. acknowledges a Swarnajayanti Fellowship from the Department of Science and Technology, India (DST/SJF/PSA-03/2016-17) and thanks the Humboldt Foundation for support that enabled his sabbatical at MPA Garching. We thank Michael McDonald for useful discussions and providing {\it Phoenix} mass profile.

\bibliographystyle{mnras}
\bibliography{references}

\begin{thebibliography}{}
\makeatletter
\relax
\def\mn@urlcharsother{\let\do\@makeother \do\$\do\&\do\#\do\^\do\_\do\%\do\~}
\def\mn@doi{\begingroup\mn@urlcharsother \@ifnextchar [ {\mn@doi@}
  {\mn@doi@[]}}
\def\mn@doi@[#1]#2{\def\@tempa{#1}\ifx\@tempa\@empty \href
  {http://dx.doi.org/#2} {doi:#2}\else \href {http://dx.doi.org/#2} {#1}\fi
  \endgroup}
\def\mn@eprint#1#2{\mn@eprint@#1:#2::\@nil}
\def\mn@eprint@arXiv#1{\href {http://arxiv.org/abs/#1} {{\tt arXiv:#1}}}
\def\mn@eprint@dblp#1{\href {http://dblp.uni-trier.de/rec/bibtex/#1.xml}
  {dblp:#1}}
\def\mn@eprint@#1:#2:#3:#4\@nil{\def\@tempa {#1}\def\@tempb {#2}\def\@tempc
  {#3}\ifx \@tempc \@empty \let \@tempc \@tempb \let \@tempb \@tempa \fi \ifx
  \@tempb \@empty \def\@tempb {arXiv}\fi \@ifundefined
  {mn@eprint@\@tempb}{\@tempb:\@tempc}{\expandafter \expandafter \csname
  mn@eprint@\@tempb\endcsname \expandafter{\@tempc}}}

\bibitem[\protect\citeauthoryear{{Akahori} et~al.,}{{Akahori}
  et~al.}{2020}]{akahouri2020}
{Akahori} T.,  et~al., 2020, arXiv e-prints, \href
  {https://ui.adsabs.harvard.edu/abs/2020arXiv200405724A} {p. arXiv:2004.05724}

\bibitem[\protect\citeauthoryear{{Babyk}, {McNamara}, {Nulsen}, {Russell},
  {Vantyghem}, {Hogan}  \& {Pulido}}{{Babyk} et~al.}{2018}]{babyk2018}
{Babyk} I.~V.,  {McNamara} B.~R.,  {Nulsen} P.~E.~J.,  {Russell} H.~R.,
  {Vantyghem} A.~N.,  {Hogan} M.~T.,   {Pulido} F.~A.,  2018, \mn@doi [\apj]
  {10.3847/1538-4357/aacce5}, \href
  {https://ui.adsabs.harvard.edu/abs/2018ApJ...862...39B} {862, 39}

\bibitem[\protect\citeauthoryear{{Bennert}, {Auger}, {Treu}, {Woo}  \&
  {Malkan}}{{Bennert} et~al.}{2011}]{bennert11}
{Bennert} V.~N.,  {Auger} M.~W.,  {Treu} T.,  {Woo} J.-H.,   {Malkan} M.~A.,
  2011, \mn@doi [\apj] {10.1088/0004-637X/726/2/59}, \href
  {https://ui.adsabs.harvard.edu/abs/2011ApJ...726...59B} {726, 59}

\bibitem[\protect\citeauthoryear{{Bennett} et~al.,}{{Bennett}
  et~al.}{2013}]{Bennett13}
{Bennett} C.~L.,  et~al., 2013, \mn@doi [\apjs] {10.1088/0067-0049/208/2/20},
  \href {https://ui.adsabs.harvard.edu/abs/2013ApJS..208...20B} {208, 20}

\bibitem[\protect\citeauthoryear{{Choudhury}, {Sharma}  \&
  {Quataert}}{{Choudhury} et~al.}{2019}]{Choudhury2019}
{Choudhury} P.~P.,  {Sharma} P.,   {Quataert} E.,  2019, \mn@doi [\mnras]
  {10.1093/mnras/stz1857}, \href
  {https://ui.adsabs.harvard.edu/abs/2019MNRAS.488.3195C} {488, 3195}

\bibitem[\protect\citeauthoryear{{Combes}}{{Combes}}{2017}]{combes17}
{Combes} F.,  2017, \mn@doi [Frontiers in Astronomy and Space Sciences]
  {10.3389/fspas.2017.00010}, \href
  {https://ui.adsabs.harvard.edu/abs/2017FrASS...4...10C} {4, 10}

\bibitem[\protect\citeauthoryear{{Hogan} et~al.,}{{Hogan}
  et~al.}{2017}]{Hogan2017}
{Hogan} M.~T.,  et~al., 2017, \mn@doi [\apj] {10.3847/1538-4357/aa9af3}, \href
  {http://adsabs.harvard.edu/abs/2017ApJ...851...66H} {851, 66}

\bibitem[\protect\citeauthoryear{{Kaiser}}{{Kaiser}}{1986}]{Kaiser86}
{Kaiser} N.,  1986, \mn@doi [\mnras] {10.1093/mnras/222.2.323}, \href
  {https://ui.adsabs.harvard.edu/abs/1986MNRAS.222..323K} {222, 323}

\bibitem[\protect\citeauthoryear{{Li}, {Bryan}, {Ruszkowski}, {Voit}, {O'Shea}
  \& {Donahue}}{{Li} et~al.}{2015}]{li15}
{Li} Y.,  {Bryan} G.~L.,  {Ruszkowski} M.,  {Voit} G.~M.,  {O'Shea} B.~W.,
  {Donahue} M.,  2015, \mn@doi [ApJ] {10.1088/0004-637X/811/2/73}, \href
  {https://ui.adsabs.harvard.edu/abs/2015ApJ...811...73L} {811, 73}

\bibitem[\protect\citeauthoryear{{McCarthy}, {Babul}, {Bower}  \&
  {Balogh}}{{McCarthy} et~al.}{2008}]{mcc08}
{McCarthy} I.~G.,  {Babul} A.,  {Bower} R.~G.,   {Balogh} M.~L.,  2008, \mn@doi
  [\mnras] {10.1111/j.1365-2966.2008.13141.x}, \href
  {https://ui.adsabs.harvard.edu/abs/2008MNRAS.386.1309M} {386, 1309}

\bibitem[\protect\citeauthoryear{{McCourt}, {Sharma}, {Quataert}  \&
  {Parrish}}{{McCourt} et~al.}{2012}]{mccourt12}
{McCourt} M.,  {Sharma} P.,  {Quataert} E.,   {Parrish} I.~J.,  2012, \mn@doi
  [\mnras] {10.1111/j.1365-2966.2011.19972.x}, \href
  {http://adsabs.harvard.edu/abs/2012MNRAS.419.3319M} {419, 3319}

\bibitem[\protect\citeauthoryear{{McDonald} et~al.,}{{McDonald}
  et~al.}{2012}]{McDonald2012}
{McDonald} M.,  et~al., 2012, \mn@doi [\nat] {10.1038/nature11379}, \href
  {https://ui.adsabs.harvard.edu/abs/2012Natur.488..349M} {488, 349}

\bibitem[\protect\citeauthoryear{{McDonald} et~al.,}{{McDonald}
  et~al.}{2014}]{McDonald2014}
{McDonald} M.,  et~al., 2014, \mn@doi [\apj] {10.1088/0004-637X/784/1/18},
  \href {https://ui.adsabs.harvard.edu/abs/2014ApJ...784...18M} {784, 18}

\bibitem[\protect\citeauthoryear{{McDonald} et~al.,}{{McDonald}
  et~al.}{2019a}]{McDonald2019}
{McDonald} M.,  et~al., 2019a, arXiv e-prints, \href
  {http://adsabs.harvard.edu/abs/2019arXiv190408942M} {}

\bibitem[\protect\citeauthoryear{{McDonald} et~al.,}{{McDonald}
  et~al.}{2019b}]{McDonald2019a}
{McDonald} M.,  et~al., 2019b, \mn@doi [\apj] {10.3847/1538-4357/aaf394}, \href
  {https://ui.adsabs.harvard.edu/abs/2019ApJ...870...85M} {870, 85}

\bibitem[\protect\citeauthoryear{{O'Sullivan} et~al.,}{{O'Sullivan}
  et~al.}{2012}]{ewan2012}
{O'Sullivan} E.,  et~al., 2012, \mn@doi [\mnras]
  {10.1111/j.1365-2966.2012.21459.x}, \href
  {https://ui.adsabs.harvard.edu/abs/2012MNRAS.424.2971O} {424, 2971}

\bibitem[\protect\citeauthoryear{{Prasad}, {Sharma}  \& {Babul}}{{Prasad}
  et~al.}{2015}]{Prasad15}
{Prasad} D.,  {Sharma} P.,   {Babul} A.,  2015, \mn@doi [ApJ]
  {10.1088/0004-637X/811/2/108}, \href
  {https://ui.adsabs.harvard.edu/abs/2015ApJ...811..108P} {811, 108}

\bibitem[\protect\citeauthoryear{{Prasad}, {Sharma}  \& {Babul}}{{Prasad}
  et~al.}{2018}]{Prasad2018}
{Prasad} D.,  {Sharma} P.,   {Babul} A.,  2018, \mn@doi [\apj]
  {10.3847/1538-4357/aacce8}, \href
  {https://ui.adsabs.harvard.edu/abs/2018ApJ...863...62P} {863, 62}

\bibitem[\protect\citeauthoryear{{Pulido} et~al.,}{{Pulido}
  et~al.}{2018}]{Pulido2018}
{Pulido} F.~A.,  et~al., 2018, \mn@doi [\apj] {10.3847/1538-4357/aaa54b}, \href
  {https://ui.adsabs.harvard.edu/abs/2018ApJ...853..177P} {853, 177}

\bibitem[\protect\citeauthoryear{{Russell} et~al.,}{{Russell}
  et~al.}{2017}]{Russell2017}
{Russell} H.~R.,  et~al., 2017, \mn@doi [\apj] {10.3847/1538-4357/836/1/130},
  \href {https://ui.adsabs.harvard.edu/abs/2017ApJ...836..130R} {836, 130}

\bibitem[\protect\citeauthoryear{{Sharma}, {McCourt}, {Quataert}  \&
  {Parrish}}{{Sharma} et~al.}{2012}]{sharma12}
{Sharma} P.,  {McCourt} M.,  {Quataert} E.,   {Parrish} I.~J.,  2012, \mn@doi
  [MNRAS] {10.1111/j.1365-2966.2011.20246.x}, \href
  {https://ui.adsabs.harvard.edu/abs/2012MNRAS.420.3174S} {420, 3174}

\bibitem[\protect\citeauthoryear{{Stern}, {Fielding}, {Faucher-Gigu{\`e}re}  \&
  {Quataert}}{{Stern} et~al.}{2019}]{Stern2019}
{Stern} J.,  {Fielding} D.,  {Faucher-Gigu{\`e}re} C.-A.,   {Quataert} E.,
  2019, arXiv e-prints, \href
  {https://ui.adsabs.harvard.edu/abs/2019arXiv190607737S} {}

\bibitem[\protect\citeauthoryear{{Ueda}, {Hayashida}, {Anabuki}, {Nakajima},
  {Koyama}  \& {Tsunemi}}{{Ueda} et~al.}{2013}]{ueda2013}
{Ueda} S.,  {Hayashida} K.,  {Anabuki} N.,  {Nakajima} H.,  {Koyama} K.,
  {Tsunemi} H.,  2013, \mn@doi [\apj] {10.1088/0004-637X/778/1/33}, \href
  {https://ui.adsabs.harvard.edu/abs/2013ApJ...778...33U} {778, 33}

\bibitem[\protect\citeauthoryear{{Voit}, {Kay}  \& {Bryan}}{{Voit}
  et~al.}{2005}]{voit05}
{Voit} G.~M.,  {Kay} S.~T.,   {Bryan} G.~L.,  2005, \mn@doi [\mnras]
  {10.1111/j.1365-2966.2005.09621.x}, \href
  {https://ui.adsabs.harvard.edu/abs/2005MNRAS.364..909V} {364, 909}

\bibitem[\protect\citeauthoryear{{Voit}, {Donahue}, {Bryan}  \&
  {McDonald}}{{Voit} et~al.}{2015a}]{voit15N}
{Voit} G.~M.,  {Donahue} M.,  {Bryan} G.~L.,   {McDonald} M.,  2015a, \mn@doi
  [Nature] {10.1038/nature14167}, \href
  {https://ui.adsabs.harvard.edu/abs/2015Natur.519..203V} {519, 203}

\bibitem[\protect\citeauthoryear{{Voit}, {Donahue}, {O'Shea}, {Bryan}, {Sun}
  \& {Werner}}{{Voit} et~al.}{2015b}]{voit15E}
{Voit} G.~M.,  {Donahue} M.,  {O'Shea} B.~W.,  {Bryan} G.~L.,  {Sun} M.,
  {Werner} N.,  2015b, \mn@doi [\apjl] {10.1088/2041-8205/803/2/L21}, \href
  {https://ui.adsabs.harvard.edu/abs/2015ApJ...803L..21V} {803, L21}

\makeatother
\end{thebibliography}
\label{lastpage}
\end{document}